\documentclass[proceedings]{JHEP3}
\PrHEP{ hep2001}
\usepackage{epsfig,multicol,amsmath}                    

\conference{International Europhysics Conference on HEP} 

\title{\boldmath Jets and Multiplicities in $e^+e^-$}

\author{\speaker{Joost Vossebeld}\\
        CERN, CH-1211 Geneva 23, Switzerland\\
        E-mail: \email{Joost.Vossebeld@cern.ch}}

\newcommand{\epem} {$e^+e^-$}
\newcommand{\as} {$\alpha_s$}
\newcommand{\ca} {$C_A$}
\newcommand{\cf} {$C_F$}
\newcommand{\ycut}   {\ensuremath{y_{\mathrm{cut}}}}

\newcommand{\rfour}               {\ensuremath{R_4}}

\newcommand{\dtwo}               {\ensuremath{D_2}}

\abstract{Recent results on jet and inclusive charged particle production 
in hadronic \epem\ interactions are reviewed.}

\begin{document}

\section{Introduction}
The production of jets and hadrons in \epem\ interactions
provides a suitable laboratory to study quantum chromodynamics. Both the 
coupling strength and the group structure of the theory can be determined.
We discuss a number of recent measurements concerning 
jet and hadron production. These include a determination of the strong 
coupling constant, \as, from 4-jet rates\cite{r4delphi}, two  
studies of 4-jet angular correlations\cite{cfopal,cfaleph}, in which 
\as\ and the colour factors, $C_A$ and $C_F$, are determined, a study of 
charged particle multiplicity in 3-jet events\cite{delphi3jetmult} and 
two helicity analyses of charged hadron 
production\cite{tljade,tldelphi}.

%
%
  \section{\boldmath Determination of \as\ from 4-jet rates}

In a recent study\cite{r4delphi} DELPHI has measured $n$-jet rates, $R_n$, 
as a function of the jet resolution parameter \ycut, at
centre-of-mass energies in the range 89-207~GeV, using various 
jet algorithms. 
The 4-jet rate, \rfour, is compared to NLO QCD 
predictions\cite{debrecen1} to determine \as. 
As there are expected to be considerable higher order contributions, still 
missing in these predictions, in particular due to large 
logarithmic terms, DELPHI uses the method of 
scale optimisation when determining \as. This 
implies that both \as\ and the renormalisation scale parameter 
$x_\mu={\mu_R^2\over Q^2}$
are varied when fitting the NLO predictions to the data.
It is argued that the obtained optimal scale, $x_\mu^{\rm opt}$, accounts 
for missing higher order contributions.

\EPSFIGURE{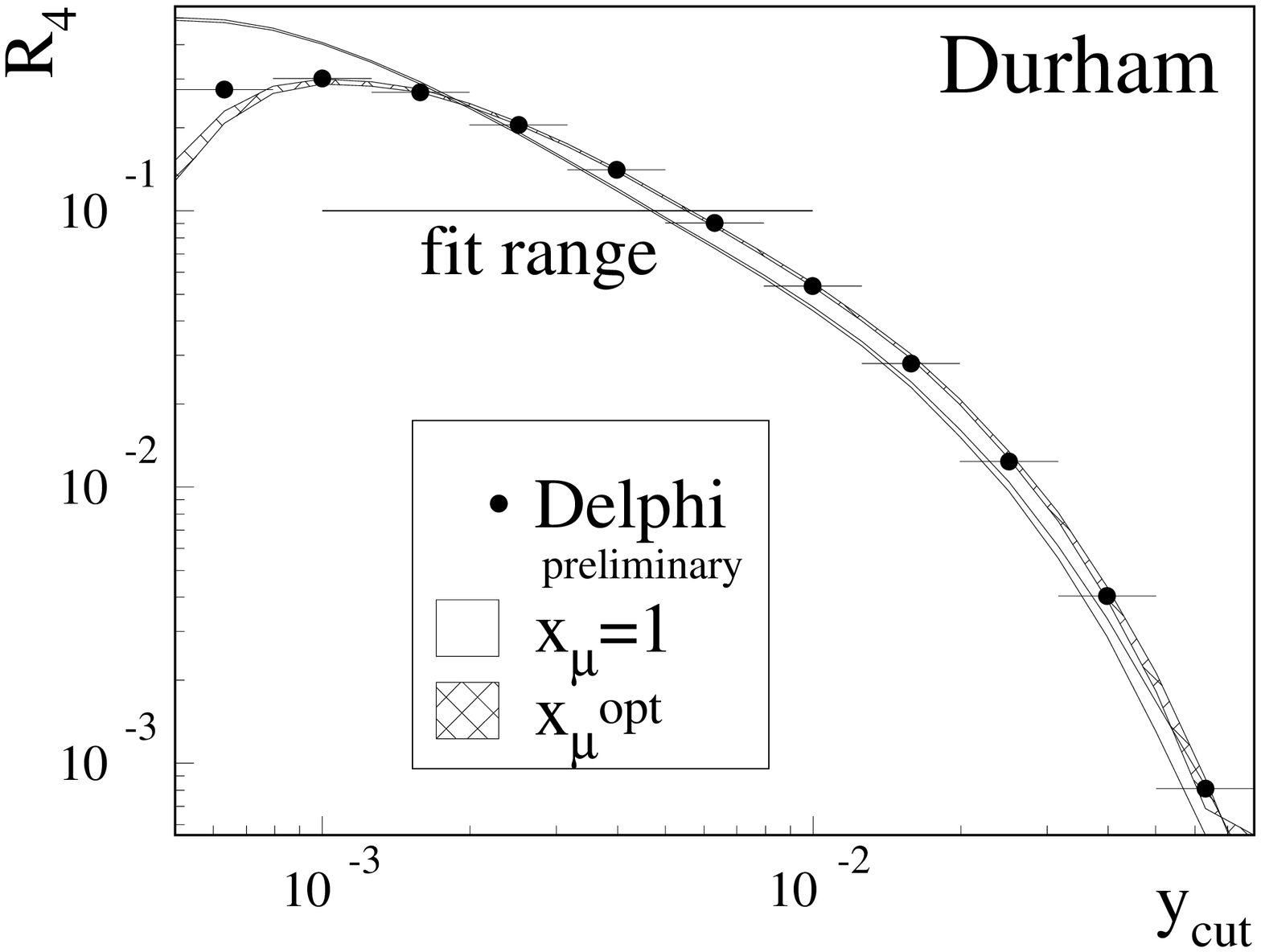,width=0.38\textwidth}{Durham \rfour\ distribution at $\sqrt{s}=91$~GeV compared to fitted NLO predictions with a fixed or a fitted scale. \label{fig1}}

Fig.~\ref{fig1} shows the result for \rfour\ based on the Durham\cite{durham} 
jet algorithm, 
at $\sqrt{s}=91$~GeV, compared to fitted theory 
predictions.
The fitted value for the strong coupling constant is: $\alpha_s(M_Z) = 0.1178 \pm 0.0012 ({\rm exp.}) \pm 0.0023 ({\rm had.}) \pm 0.0014 ({\rm scale})$ 
and the corresponding optimal scale: $x_\mu^{\rm opt}=0.015$. The low 
value obtained for the scale suggests that missing higher orders 
are indeed important. The scale uncertainty on \as\ is determined 
by varying $x_\mu$ between ${1\over 2}x_\mu^{\rm opt}$ and 
$2 x_\mu^{\rm opt}$, yielding a smaller uncertainty than obtained 
with the conventional variation of 
$x_\mu$ between $1\over 4$ and $4$. The obtained \as\ value is in 
good agreement with a similar ALEPH\cite{alephr4} result, obtained 
using resummed NLO predictions.

Using \rfour\ measurements from different centre-of-mass energies, DELPHI
has also studied the scale dependence of \as, finding good agreement with the 
running behaviour predicted in QCD.

%
%
  \section{\boldmath Determination of \as, \ca\ and \cf\ from 4-jet angular 
correlations}

OPAL\cite{cfopal} and ALEPH\cite{cfaleph} have 
recently presented studies of angular correlations in 4-jet events 
at $\sqrt{s}=91$~GeV.
A combined fit of theory predictions to these correlations and to jet rates 
is used to determine \as, $C_A$ and $C_F$. The theory 
predictions\cite{debrecen1,debrecen2} compared to 
are to NLO accuracy for the angular correlations and to resummed NLO accuracy 
for the jet rates.

In both the OPAL and the ALEPH studies the 
theory predictions have been fit\-ted 
simultaneously to 4 angular variables, to \rfour\ and, 
in the OPAL study, also to the differential 2-jet rate, \dtwo.
The 4-jet angles measured are the Bengtsson-Zerwas\cite{BZ} angle, 
$\chi _{BZ}$, the modified Nachtmann-Reiter\cite{NR} angle, $\Theta _{NR}$, 
the K\" orner-Schierholtz-Willrodt\cite{KSW} angle, $\Phi _{KSW}$, and
the angle between the two lowest energy jets, $\alpha_{34}$.
Fig.~\ref{fig2} shows these angles as measured by OPAL, compared to the 
fitted NLO predictions.

\DOUBLEFIGURE[b]{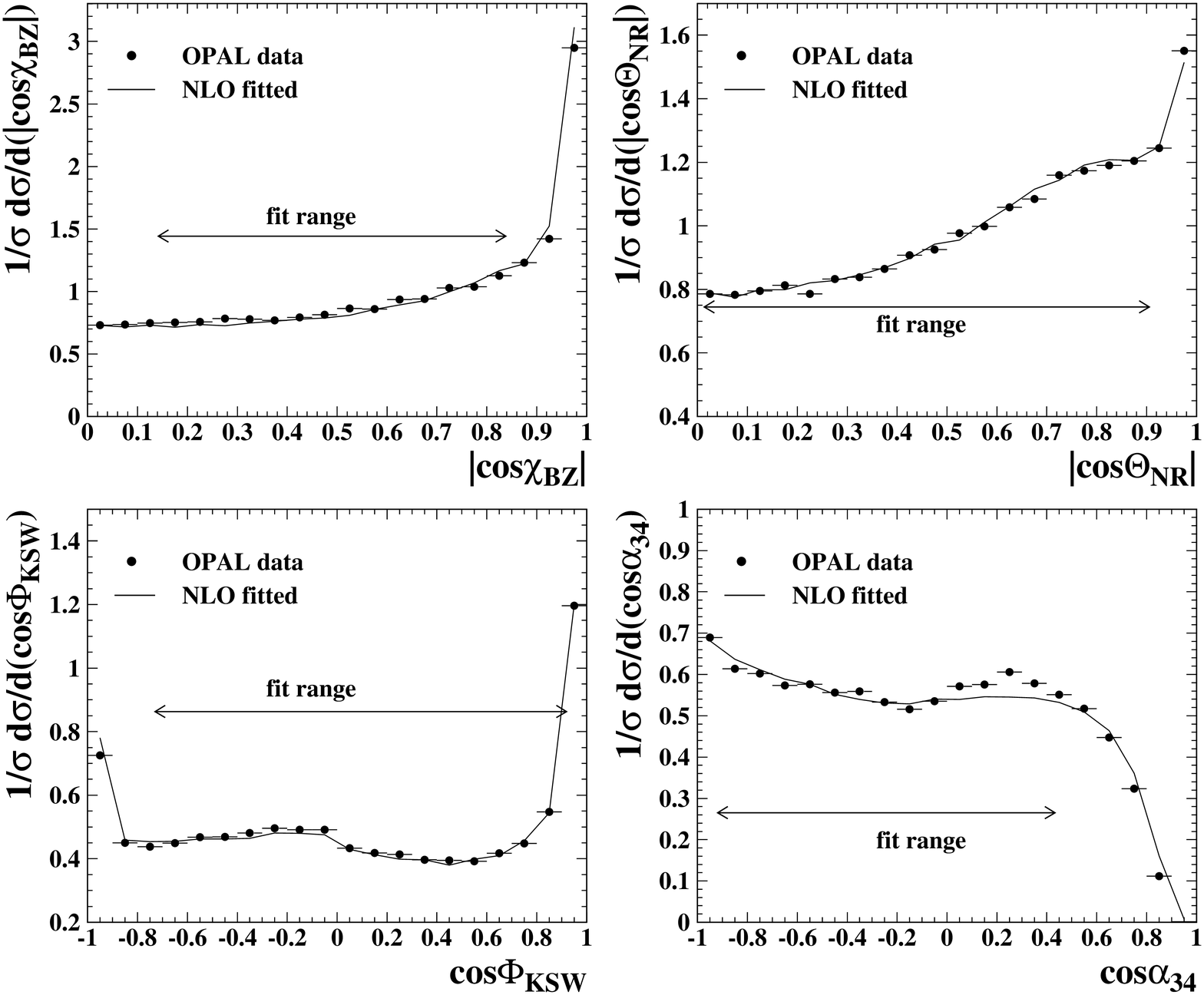,width=0.45\textwidth}{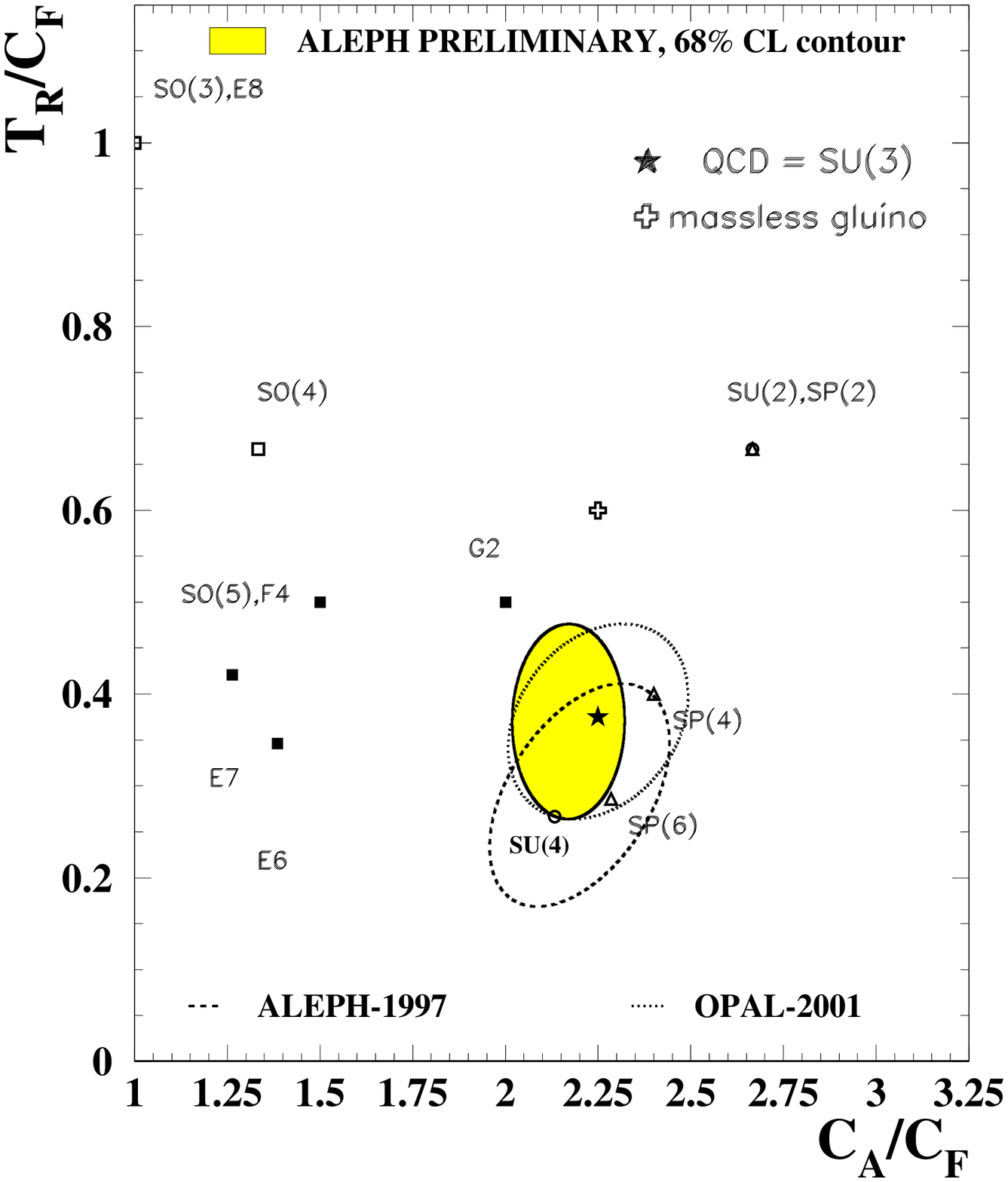,width=0.35\textwidth}{4-jet angles measured by OPAL at $\sqrt{s}=91$~GeV, compared to fitted NLO predictions. \label{fig2}}{68\% confidence level contours in $\left({T_R\over C_F},{C_A\over C_F}\right)$, compared to the expectated values in various Lie group structures. \label{fig3}}

The fit results obtained by OPAL are: $\alpha_s(M_Z)=0.120\pm0.011\pm0.020$, 
$C_A=3.02\pm0.25\pm0.49$, $C_F=1.34\pm0.13\pm0.22$, and by
ALEPH: 
$\alpha_s(M_Z)=0.119\pm0.006\pm0.022$, $C_A=2.93\pm0.14\pm0.49$, 
$C_F=1.35\pm0.07\pm0.22$, where the uncertainties are the statistical and 
systematic uncertainties, respectively.
In Fig.~\ref{fig3} 68\% confidence 
level contours in the $\left({T_R\over C_F},{C_A\over C_F}\right)$ plane 
are shown from the results presented here and from an earlier ALEPH study. 
The results are compared to the expected values for different group 
structures. All results agree 
well with the SU(3) group structure while several alternatives are 
disfavoured.

%
%

  \section{Charged particle multiplicity in 3-jet events}
Multiplicity differences in the fragmentation of quark and gluon jets are 
of great interest as they provide a direct measure of the colour factor 
ratio $C_A\over C_F$. Most measurements of the multiplicity in gluon 
jets, however, rely on a jet algorithm to separate the gluon fragmentation 
products from those of the quarks in 3-jet events, rendering the obtained 
results biased. 
Unbiased gluon jet multiplicity has so far only been measured in 
$\Upsilon$-decays\cite{gluoncleo1,gluoncleo2} and 
in \epem\ 3-jet events, 
where the gluon is associated with the highest energy jet and it's 
fragmentation products can therefore be separated in  
one hemisphere of the event\cite{gluonopal}. 

Recently a formalism\cite{eden} has been proposed (in MLLA\cite{mlla})
to express the multiplicity 
in 3-jet events as a function of the unbiased gluon jet multiplicity and a  
biased quark jet multiplicity, where the latter is defined as the 
multiplicity in hadronic \epem\ events with no gluon radiation harder 
than the scale associated with the gluon jet in the 3-jet event. 
\begin{subequations}
\begin{eqnarray} N_{q\bar{q}g}&=&N_{q\bar{q}}(L_{q\bar{q}},\kappa_{\perp Lu})+{1\over 2} N_{gg}(\kappa_{\perp Le})\label{eqa}\\
N_{q\bar{q}g}&=&N_{q\bar{q}}(L,\kappa_{\perp Lu})+{1\over 2} N_{gg}(\kappa_{\perp Lu})\label{eqb}
\end{eqnarray}
\end{subequations}
where $L={\ln}{s \over \Lambda^2}$, $L_{q\bar{q}}={\ln}{s_{q\bar{q}} \over \Lambda^2}$, $\kappa_{\perp Lu}=\ln{s_{qg}s_{\bar{q}g} \over s\Lambda^2}$, $\kappa_{\perp Le}=\ln{s_{qg}s_{\bar{q}g} \over s_{q\bar{q}}\Lambda^2}$ and $s_{ij}=(p_i+p_j)^2$. The two alternative formulations reflect an ambiguity in defining
the transverse momentum of the emitted gluon with respect to the 
$q\bar{q}$ system. 
In the same formalism both the biased quark jet multiplicity and the 
energy dependence of the gluon jet multiplicity can be derived from the 
inclusive multiplicity in hadronic \epem\ events.

\EPSFIGURE[b]{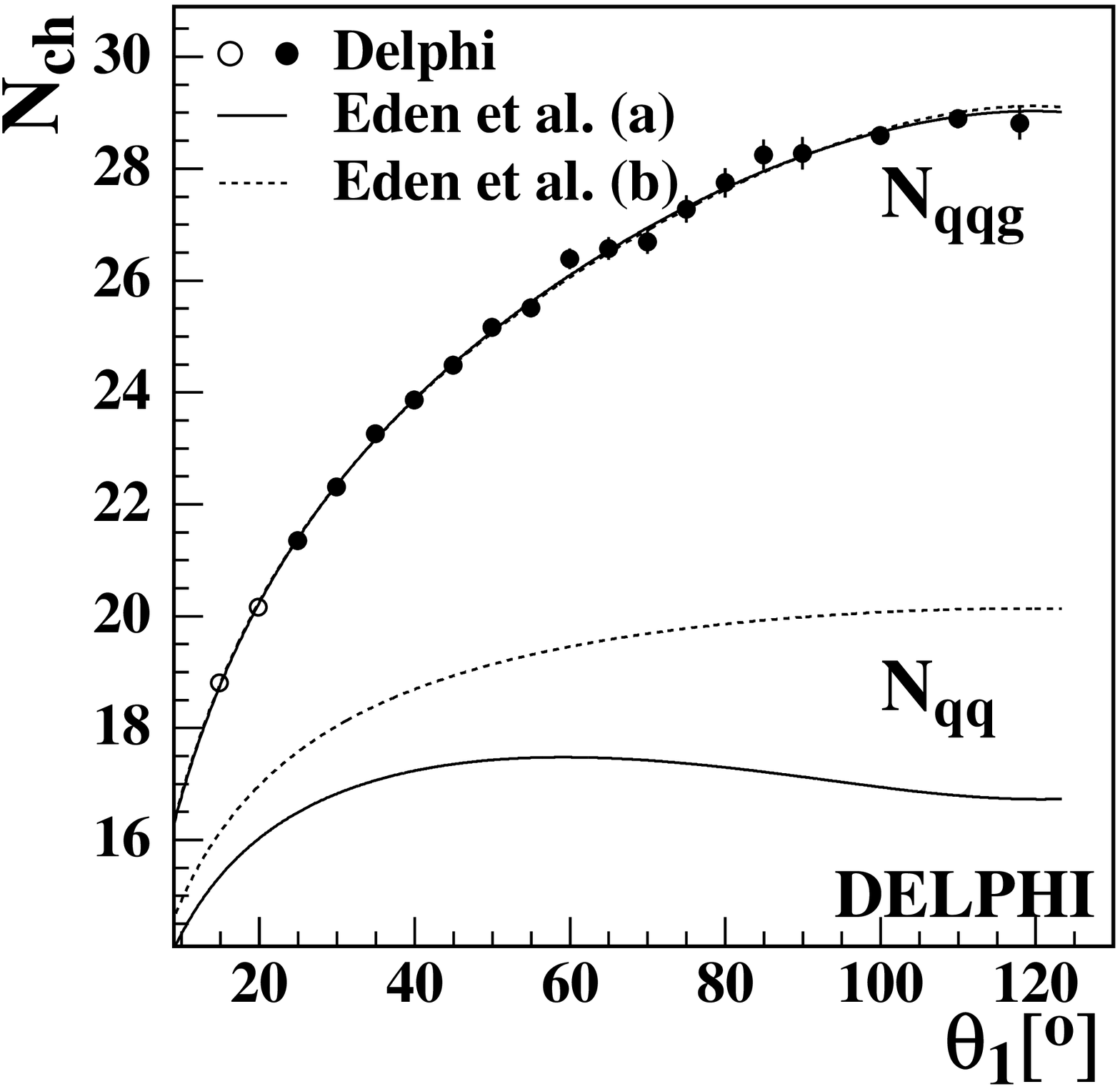,width=0.36\textwidth}{Charged particle multiplicity in 3-jet events in bins of $\theta_1$, compared to predictions based on equations \ref{eqa} and \ref{eqb}. \label{fig4}}

In a recent analysis\cite{delphi3jetmult} the DELPHI collaboration has 
measured the charged particle multiplicity in 3-jet events, in which the 
two lowest energy jets have identical angles with respect to the highest 
energy jet. In these so-called ``Y'' events all scales in equations 
\ref{eqa} and \ref{eqb} are determined by the angle between the two lowest 
energy jets, $\theta_1$. The charged multiplicity in 3-jet events as a 
function of $\theta_1$ is shown in Fig.~\ref{fig4}. The data are compared to  
predictions based on the above formalism, where the normalisation of the 
gluon jet multiplicity has been fixed to the direct measurement of \cite{gluoncleo1} and an additional offset is fitted to the data to account 
for the bias in the multiplicity due to c and b quarks which are not 
considered in the formalism.

The formalism (equation \ref{eqa}) has also been used to obtain a measurement 
of $C_A\over C_F$ which determines the ratio of the 
energy slopes of the multiplicity in gluon and quark jets. DELPHI obtains:
${C_A\over C_F}=2.221\pm0.032{\rm (stat.)}\pm0.047{\rm (exp.)}\pm0.058{\rm (had.)}\pm0.075{\rm (theo.)}$.
In agreement with the QCD expectation of 2.25. Recently DELPHI has extended 
this 
study to include also asymmetric 3-jet events, obtaining similar results.

%
%
  \section{Helicity analysis of inclusive charged hadron production}
The inclusive hadron production cross section in \epem\ annihilation 
can be expressed as a function of $x_p$, the scaled momentum of the hadrons 
and $\theta$, the angle of the hadrons with respect to the electron 
beam:
\begin{equation} \frac{{\rm d}^2\sigma^h}{{\rm d}x_p {\rm d}\cos\theta} =  \frac{3}{8}(1+\cos^2{\theta})\frac{{\rm d}\sigma^h_T}{{\rm d}x_p}+\frac{3}{4}\sin^2{\theta}\frac{{\rm d}\sigma^h_L}{{\rm d}x_p}.\end{equation}
A transverse and  a longitudinal component are distinguished, which can be separated by measuring the $\cos{\theta}$ distribution.
This separation is of interest because the relative size of the longitudinal 
component, which is associated with gluon radiation,  
provides a measure of the strong coupling constant.
A prediction up to NLO in \as\ for the longitudinal fraction has been 
given in \cite{rijken}: 
\begin{equation} \frac{\sigma_L}{\sigma_{tot}}=\frac{\alpha_s}{\pi}+\frac{\alpha_s^2}{\pi^2}\left(13.583-1.028N_f+(0.167N_f-2.750)\ln{Q^2\over \mu^2} \right)\label{eqsigl}.\end{equation}

\EPSFIGURE[b]{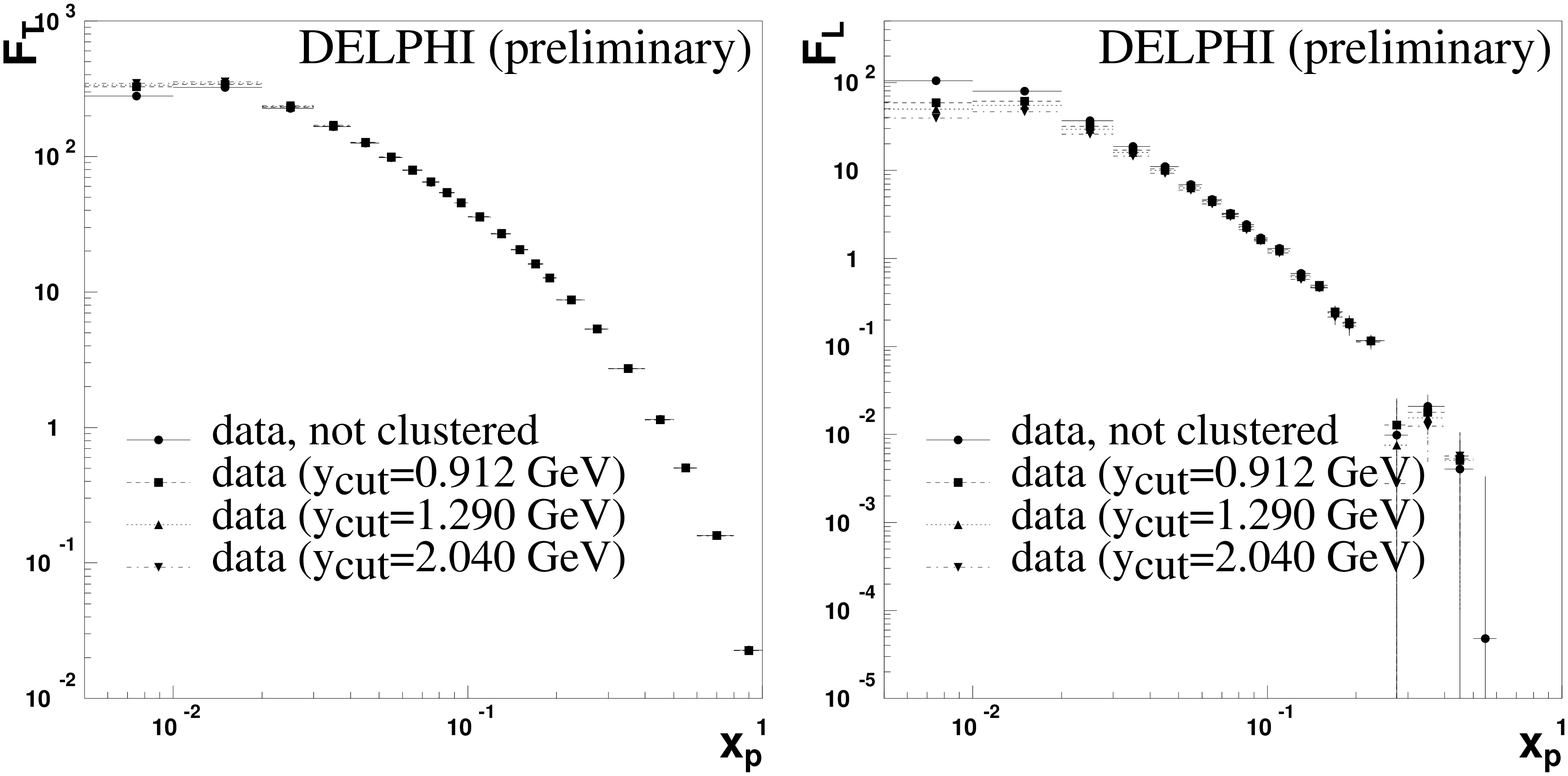,width=0.57\textwidth}{Transverse and longitudinal fragmentation functions at $\sqrt{s}=91$~GeV using angles of individual hadrons or those of jets. \label{fig5}}

In a recent DELPHI study\cite{tldelphi} of hadronic \epem\ interactions at 
$\sqrt{s}=91$~GeV,  
transverse and longitudinal charged particle fragmentation functions have been 
determined from the $\cos{\theta}$ distribution measured in bins of $x_p$. 
To study possible hadronisation effects on the hadron angles, 
which are not accounted for in equation~\ref{eqsigl}, alternative ways of 
measuring these angles were tested, either using the angles of the hadrons 
themselves or 
using the angles of the jets to which the hadrons are assigned. In the 
latter case 
jets were defined using different values of the non-scaled distance 
measure \ycut.   
The obtained transverse 
and longitudinal fragmentation functions are shown in Fig.\ref{fig5}. 
Taking the measurements with $y_{\rm cut}=1.290$~GeV as the nominal results, 
DELPHI obtains the longitudinal fraction:
$\frac{\sigma_L}{\sigma_{\rm tot}}(M_Z)=  0.0445\pm0.0006({\rm stat.})\pm0.0060({\rm syst.})$.
Using equation~\ref{eqsigl} the corresponding value for \as\ is found 
to be:
$\alpha_s(M_Z)=0.1083\pm0.0012({\rm stat.})\pm0.0119({\rm syst.})$.

In a similar analysis of data from the JADE experiment\cite{tljade} the   
$\cos{\theta}$ distribution was measured for \epem\ interactions at a 
mean centre-of-mass energy of 36.6~GeV. The longitudinal 
fraction was determined to be: 
$\frac{\sigma_L}{\sigma_{\rm tot}}(36.6~GeV)=  0.067\pm0.011({\rm stat.})\pm0.007({\rm syst.})$ and the corresponding value of the strong coupling constant:
$\alpha_s(36.6~{\rm GeV})=0.150\pm0.020({\rm stat.})\pm0.013({\rm syst.})\pm0.008({\rm scale})$.
Evolved up to $M_Z$ this corresponds to 
$\alpha_s(M_Z)=0.127^{+0.017}_{-0.018}$, 
in agreement with the DELPHI result and with the world average value 
of $\alpha_s(M_Z)$ ($0.1184\pm0.0031$\cite{bethke}).

%
%
  \section{Summary}
Various new results on jet and inclusive hadron production in \epem\ 
annihilation have been presented. The measurements have been compared to 
higher order QCD predictions to test this
theory and to extract it's coupling constant and it's colour factors. 
The obtained
results for \as\ agree with results obtained in other measurements 
and the obtained 
values for the colour factors agree with the values expected in QCD.

\end{document}